\documentclass[preprint,showpacs,preprintnumbers,amsmath,amssymb]{revtex4}
\usepackage{graphicx,epsfig,dcolumn,bm,epic,eepic,float}%
\usepackage{amsmath}
\usepackage{latexsym}
\usepackage{color}
\usepackage{cancel}
\usepackage{makeidx,shortvrb,latexsym}
\usepackage{slashed}
\begin{document}
\unitlength 1 cm
\newcommand{\nn}{\nonumber}
\newcommand{\vk}{\vec k}
\newcommand{\vp}{\vec p}
\newcommand{\vq}{\vec q}
\newcommand{\vkp}{\vec {k'}}
\newcommand{\vpp}{\vec {p'}}
\newcommand{\vqp}{\vec {q'}}
\newcommand{\bk}{{\bf k}}
\newcommand{\bp}{{\bf p}}
\newcommand{\bq}{{\bf q}}
\newcommand{\br}{{\bf r}}
\newcommand{\bR}{{\bf R}}
\newcommand{\up}{\uparrow}
\newcommand{\down}{\downarrow}
\newcommand{\fns}{\footnotesize}
\newcommand{\ns}{\normalsize}
\newcommand{\cdag}{c^{\dagger}}
\title {A detailed study of charm content of a proton in the frameworks of the
$Kimber$-$Martin$-$Ryskin$
 and  $Martin$-$Ryskin$-$Watt$   approaches}
\author{N. Olanj$^\dag$}\altaffiliation {Corresponding author, Email :   {$n\_olanj@basu.ac.ir$}
, Tel:+98-81-38381601}
\author{M. Modarres$^\ddag$}
\affiliation{$^\dag$Physics Department, Faculty of  Science, Bu-Ali
Sina University, 65178, Hamedan, Iran} 
\affiliation{$^\ddag$Physics
Department, University of  Tehran,1439955961, Tehran, Iran.}
\begin{abstract}
The charm structure function, ($F_{2}^{c\overline{c}}(x, Q^2)$), is
calculated in the framework of   the $k_t$-factorization formalism
by using the $unintegrated$ parton distribution functions ($UPDF$), which are generated through the $Kimber$ et al. ($KMR$) and $Martin$
et al. ($MRW$) procedures. The $Martin$ group ($MMHT2014$) parton
distribution functions ($PDF$) is used as the input $PDF$ for the
corresponding $UPDF$. The resulted $F_{2}^{c\overline{c}}(x, Q^2)$ is
compared with the predicted data and the calculations given by the $ZEUS$ and $H1$
collaborations, the  parton $pQCD$ theory, i.e. the general-mass
variable-flavour-number scheme ($GMVFNS$), the $LO$ collinear procedure and the saturation model
introduced by $Golec-Biernat$ and $W\ddot{u}sthoff$ ($GBW$),
respectively. In general, it is shown that the calculated charm
structure functions based on the stated above two $UPDF$  schemes are
consistent with the experimental data and other theoretical
predictions. Also, a short discussion is presented regarding the $KMR$ and $MRW$ $UPDF$ behaviors.
\end{abstract}
 \maketitle
\section{Introduction}
The heavy-quark electro-production contributes up to 25 \% to the
deep inelastic scattering ($DIS$) inclusive cross section at small $x$
\cite{1}. Therefore, the study of the heavy-quark distribution
functions of hadrons become a significant gradient  in the $Higgs$
production \cite{2} at the $LHC$ and in the $W$ and $Z$ bosons
inclusive production
 \cite{3,3p}. So  the measurement of the heavy
quarks structure functions in $DIS$ at $HERA$ is an important test
of the theory of perturbative quantum $chromodynamics$ ($pQCD$) at the small
$Bjorken$ scale \cite{3p,4} as well as the outcome of the above
$LHC$ semi-inclusive cross sections.

Usually, $pQCD$ is applied to calculate various quantities such the
hadrons structure function, the hadron-hadron differential
cross sections, etc., but, as we pointed out above, it is well known
that in the small $Bjorken$ scale region, there are some theoretical
problems \cite{m1,m2,m3}, which indicate that transverse momentum and
the reggeon play an important role \cite{m1,m2,m3}.  In recent
years, plenty of available experimental data on the various
events,
   such as the exclusive and semi-inclusive processes in the high energy collisions at $LHC$,
    indicates the necessity for computation of the transverse-dependent parton distributions,
     which are called the $unintegrated$ parton distribution functions ($UPDF$).
     Therefore, the extraction of the $UPDF$  recently become very important. The $UPDF$, $f_a(x,k_t^2,\mu^{2})$,
      are the two-scale
 dependent functions, i.e., $k_t^2$ and $\mu^{2}$, which satisfy
 the $Ciafaloni$-$Catani$-$Fiorani$-$Marchesini$ ($CCFM$)
 equations \cite{5,6,7,8}, where $x$, $ k_t$, and $\mu$ are the longitudinal momentum fraction (the $Bjorken$ variable),
  the transverse momentum, and the factorization scale, respectively. Nevertheless, solving the $CCFM$ equation is
  a mathematically involved task. Also, there is not a complete quark version of the $CCFM$ formalism.
   Therefore, to obtain the $UPDF$, the $Kimber$, $Martin$ and $Ryskin$ ($KMR$) \cite{9} and the $Martin$, $Ryskin$ and $Watt$ ($MRW$) \cite{10}
   proposed a different procedure based on the standard  $Dokshitzer$-$Gribov$-$Lipatov$-$Altarelli$-$Parisi$ ($DGLAP$) equations \cite{1a,1b,1c,1d} in the leading order ($LO$) and the
next-to-leading order ($NLO$) levels,
    respectively, along with a modification due to the angular ordering condition ($AOC$), which is the key dynamical
     property of the $CCFM$ formalism.
     \\ As evidenced by the analytic extraction of the parton distribution functions from the evolution equation (the $DGLAP$ equation) in the reference [4], the integration over the transverse momentum of the partons are performed. Therefore, the $UPDF$, $f_a$ unintegrated over the parton $k_t$, are the fundamental quantities for the phenomenological computations in the high energy collisions of hadrons.

Due to the importance of this subject, in the present work, by
studying recent data on
      charm structure functions at the small $x$, we examine the validity of
       the $KMR$ and $MRW$ approaches.

In our previous articles, we  investigated the general behavior and
stability of
 the $KMR$ and $MRW$ schemes \cite{11,12,13,14,15,16,17,18,19}. Also, we have successfully used $KMR$-$UPDF$ to calculate the inclusive production of the $W$ and $Z$ gauge vector bosons  \cite{171,181}, the semi-$NLO$ production of $Higgs$ bosons  \cite{191}, the production of forward-center and forward-forward di-jets \cite{201}, the prompt-photon pair production \cite{211} and recently in the reference \cite{221}, we explored the phenomenology of  the integral and the differential versions of the $KMR$-$UPDF$ using the angular (strong) ordering ($AOC$ ($SOC$)) constraints.
 But here, to check the reliability of generated $UPDF$,
 we  use them to calculate the observable, deep inelastic scattering the charm structure functions ($F_{2}^{c\overline{c}}(x, Q^2)$).
Then the predictions of these two methods for the charm structure
functions are compared to the  experimental measurements of
$ZEUS$ \cite{4} and $H1$ \cite{21}  as well as the parton model $pQCD$,
i.e., the general-mass variable-flavour-number scheme ($GMVFNS$)
\cite{21-1,21-2,21-3,21-4,21-5,29} and the saturation model
introduced by $Golec-Biernat$ and $W\ddot{u}sthoff$ ($GBW$)
\cite{GBW}.

The $UPDF$ are prepared using the $MMHT2014$ \cite{22} set of parton distribution function ($PDF$) in the $LO$ and $NLO$ levels.

So, the paper is organized as follows: in the section $II$, a
glimpse of the $KMR$ and $MRW$ approaches for the calculation of the
double-scale $UPDF$ is presented. The formulation of
$F_{2}^{c\overline{c}}(x, Q^2)$
 based on the $k_t$-factorization (\cite{new1,new2,new3,new4,new5}) scheme is given in the section $III$.
Finally, the results ,as well as our discussions and conclusions, are
given in the section $IV$.
\section{A glimpse of the $KMR$ and $MRW$ approaches}
The $KMR$ \cite{9} and $MRW$ \cite{10} formalisms were developed
to generate the $UPDF$,
 $f_a(x,k_t^2,\mu^{2})$, by using the given $PDF$, ($a(x,\mu^2)$ = $ xq(x,\mu^2)$ and $xg(x,\mu^2)$),
 and the corresponding splitting functions $P_{aa^\prime}(x)$ at the  $LO$ and $NLO$  levels, respectively, such that the equation (\ref{eq:new})  \cite{9,new6} is satisfied:
 \begin{eqnarray}
a(x,Q^2)=\int_0^{Q^2}\frac{d{k_t}^2}{{k_t}^2}f_{a}(x,k_{t}^2,Q^{2}),
 \label{eq:new}
\end{eqnarray}
since the $UPDF$ can only defined in the perturbative regime $k_t>Q_0$,  the integral in the equation (\ref{eq:new}) will be from $Q_0^2$ to $Q^2$, and therefore the left-hand side of this equation is converted to $a(x,Q^2)-a(x,Q_0^2)$.
 \\ It should be noted that the $k_t$-integrals of the $UPDF$ (the $f_q$ and $f_g$ functions)  obtained by the $KMR$ and the $MRW$ 
approaches are only approximately equal to the integrated ordinary $PDF$ ($q$ and $g$) that come from a global fit to data using
 the conventional collinear approximation. As it is stated in the reference \cite{10}, the two sides of the equation (\ref{eq:new}) are mathematically equivalent as far as we neglect the singularity of the splitting functions, $P_{qq}(z)$ and $P{gg}(z)$, at $z = 1$, corresponding to the soft gluon emission. Otherwise, by considering the singularity of the splitting functions and consequently the cutoff, according to the reference \cite{new7}, the difference between the two sides of the equation (\ref{eq:new}) can be eliminated by using the cutoff dependent $PDF$, that comes from a global fit to data using the $k_t$-factorization procedures, instead of the ordinary $PDF$. We show in the table 1, that the discrepancy between the $k_t$-integrals of the $KMR$-$UPDF$ (for the gluon and the up and charm quarks) with the input $MMHT2014-LO$ $PDF$ and their corresponding $MMHT2014-LO$ $PDF$ lies approximately in the uncertainty band of $MMHT2014-LO$ $PDF$ \cite{22}. According to the reference \cite{WattWZ}, the use of ordinary integrated $PDF$ is adequate for the initial investigations and descriptions of exclusive processes. Also, we have shown in the reference \cite{221} that the usual global fitted $PDF$  instead of the cutoff dependent $PDF$ can be used for generating
the $UPDF$ of  the integral version of the $KMR$ approach with the $AOC$ constraint with good approximation.
\\ These two procedures, which are reviewed
  in this section, are the modifications to the standard $DGLAP$ evolution equations by  imposing
  the $AOC$, which is the consequence of the coherent gluon emissions.

  Therefore, in the $KMR$ approach, the separation of the real and virtual contributions in the $DGLAP$ evolution chain at the $LO$ level,
 leads to the following forms for the quark and gluon $UPDF$:
\begin{eqnarray}
f_{q}(x,k_{t}^2,\mu^{2})&=&T_q(k_t,\mu)\frac{\alpha_s({k_t}^2)}{2\pi}
\nonumber\\&\times&
\int_x^{1-\Delta}dz\Bigg[P_{qq}(z)\frac{x}{z}\,q\left(\frac{x}{z} ,
{k_t}^2 \right)\cr &+& P_{qg}(z)\frac{x}{z}\,g\left(\frac{x}{z} ,
{k_t}^2 \right)\Bigg],
 \label{eq:8}
\end{eqnarray}
\begin{eqnarray}
f_{g}(x,k_{t}^2,\mu^{2})&=&T_g(k_t,\mu)\frac{\alpha_s({k_t}^2)}{2\pi}
\nonumber\\&\times& \int_x^{1-\Delta}dz\Bigg[\sum_q
P_{gq}(z)\frac{x}{z}\,q\left(\frac{x}{z} , {k_t}^2 \right) \cr &+&
P_{gg}(z)\frac{x}{z}\,g\left(\frac{x}{z} , {k_t}^2 \right)\Bigg],
 \label{eq:9}
\end{eqnarray}
respectively, where $P_{aa^{\prime}}(x)$ are the corresponding
splitting functions, and the survival probability factors, $T_a$,
are evaluated from:
\begin{eqnarray}
T_a(k_t,\mu)&=&\exp\Bigg[-\int_{k_t^2}^{\mu^2}\frac{\alpha_s({k'_t}^2)}{2\pi}\frac{{dk'_t}^{2}}{{k'_t}^{2}}
\cr &\times& \sum_{a'}\int_0^{1-\Delta}dz'P_{a'a}(z')\Bigg].
 \label{eq:5}
\end{eqnarray}
The $AOC$ on the last step of the evolutionary process determined
the cutoff, $\Delta=1-z_{max}=\frac{k_t}{\mu+k_t}$, to prevent
$z=1$ singularities in the splitting functions \cite{9}, which
arises from the soft gluon emission. In the $KMR$ approach, $T_a$ is
considered to be unity for $k_t>\mu$. This constraint and its
 interpretation in terms of the strong ordering condition gives the $KMR$ approach a smooth
  behavior over the small-$x$ region, which is generally governed by the
   $Balitski-Fadin- Kuraev-Lipatov$ ($BFKL$) evolution equation \cite{23,24}.

In the $MRW$ approach \cite{10}, the same separation of real and
virtual contributions to
  the $DGLAP$ evolution is done, but the procedure is at the $NLO$ level, i.e.,
\begin{eqnarray}
f_{a}(x,k_{t}^2,\mu^{2})&=&\int_x^{1}dz
T_a(k^{2},\mu^{2})\frac{\alpha_s({k}^2)}{2\pi} \nonumber\\&\times&
\sum_{b=q,g}P_{ab}^{(0+1)}(z)\,b\left(\frac{x}{z} , {k}^2
\right)\Theta(\mu^{2}-k^{2}),
 \label{eq:10}
\end{eqnarray}
where
\begin{eqnarray}
P_{ab}^{(0+1)}(z)&=&P_{ab}^{(0)}(z)+\frac{\alpha_s}{2\pi}P_{ab}^{(1)}(z),\nonumber\\k^2&=&\frac{k_t^2}{1-z}
 \label{eq:11},
\end{eqnarray}
and
\begin{eqnarray}
T_a(k^{2},\mu^{2})&=&\exp\Bigg(-\int_{k^2}^{\mu^2}\frac{\alpha_s({\kappa}^2)}{2\pi}\frac{{d\kappa}^{2}}{{\kappa}^{2}}
\cr &\times& \sum_{b=q,g}\int_0^{1}d\zeta \zeta
P_{ba}^{(0+1)}(\zeta)\Bigg).
 \label{eq:12}
\end{eqnarray}
 $P_{ab}^{(0)}$ and   $P_{ab}^{(1)}$   functions in the above
 equations
correspond to the $LO$ and $NLO$ contributions of the splitting
functions, which are given in the reference \cite{25},
respectively.

It is evident from the equation (\ref{eq:10}) that, in the $MRW$
approach, the $UPDF$ are defined such that, to ensure $k^{2}<\mu^2$.
Therefore, the $MRW$ approach is more in compliance with the $DGLAP$
evolution equations requisites, unlike the $KMR$ approach that the
$AOC$ spreads the $UPDF$ to whole transverse
momentum region, and it makes the results sum up the leading $DGLAP$
and $BFKL$ logarithms. Unlike the $KMR$ approach, where the $AOC$ is imposed on the all of the terms of the equations
(\ref{eq:8}) and (\ref{eq:9}), in the $MRW$ approach, the $AOC$ is imposed by the terms in which the splitting functions
are singular, i.e., the terms which include $P_{qq}$ and $P_{gg}$.  
\section{The formulation of $F_{2}^{c\overline{c}}(x, Q^2)$ in the $k_t$-factorization approach }
Here we briefly describe the different steps for calculations of the
charm structure functions, $F_{2}^{c\overline{c}}(x, Q^2)$, in the
$k_t$-factorization approach \cite{26}. Since the gluons in the
proton can only contribute to $F_{2}^{c\overline{c}}(x, Q^2)$ through
the intermediate quark,
 so one should calculate the charm structure functions in the $k_t$-factorization approach by
using the gluons and quarks $UPDF$. In this level, there are six
diagrams corresponding to the subprocess $g \rightarrow q
\overline{q}$ and $q \rightarrow qg$, (see the figure 6 of the
reference  \cite{27}). Following these six diagrams \cite{27}, by
considering a physical gauge for the gluon, i.e.,
$A^{\mu}q^{\prime}_{\mu}=0$ $(q^\prime=q+xp)$, only the ladder-type
diagrams (for example the quark box and the crossed box
approximations to the photon-gluon subprocess) remain valid for the
calculation (see the  figure  7 of the reference \cite{14}). These
contributions may be written in the $k_t$-factorization form, by
using the $UPDF$ which are generated
through the $KMR$ and $MRW$ formalisms, as follows: \\(i) For the
gluons,
\begin{eqnarray}
{F_2}^{c\overline{c}}_{g \rightarrow q \overline{q}}(x,Q^2) &=&
e_c^2 \frac{Q^2}{4\pi} \int\frac{dk_t^2}{k_t^4}\int_0^{1}d\beta\int
d^2\kappa_t \alpha_s(\mu^2) f_g\left(\frac{x}{z},k_t^2,\mu^2\right)
 \Theta(1-\frac{x}{z})\nonumber\\
\Bigg \lbrace  [\beta^2 &+& (1-\beta^2)] (
\frac{\bf{\kappa_t}}{D_1}-\frac{(\bf{\kappa_t}-\bf{k_t})}{D_2} )^2 +
[m_c^2+4Q^2\beta^2(1-\beta)^2] (\frac{1}{D_1}-\frac{1}{D_2})^2 \Bigg
\rbrace
 \label{eq:2},
\end{eqnarray}
where, in the above equation, in which the graphical representations
of $k_t$ and $\kappa_t$  are introduced in the
 figure 7 of the reference \cite{14}, the variable $\beta$ is defined as the light-cone fraction
  of the photon momentum carried by the internal quark \cite{9}. Also, the denominator factors are
\begin{eqnarray}
D_1&=&\kappa_t^2+\beta(1-\beta)Q^2+m_c^2,\nonumber\\D_2&=&({\bf{\kappa_t}}
-{\bf{k_t}})^2 +\beta(1-\beta)Q^2+m_c^2
 \label{eq:3},
\end{eqnarray}
and
\begin{eqnarray}
\frac{1}{z}=1+\frac{\kappa_{t}^{2}+m_c^2}{(1-\beta)Q^2}+\frac{k_t^2+\kappa_t^2-2{\bf{\kappa_t}}.{\bf{k_t}}+m_c^2}{\beta
Q^2}
 \label{eq:a}.
\end{eqnarray}
As in the reference \cite{28}, the scale $\mu$ controls both the
unintegrated partons and the $QCD$ coupling constant ($\alpha_s$),
and in the former case, it is chosen as follows,
\begin{eqnarray}
\mu^2=k_t^2+\kappa_t^2+m_q^2
 \label{eq:b}.
\end{eqnarray}
The charm quark mass is taken to be $m_c = 1.27GeV$. \\(ii) For
the quarks,
\begin{eqnarray}
{F_2}^{c\overline{c}}_{q \rightarrow qg}(x,Q^2)= &e_c^2&
\int_{k_0^2}^{Q^2}
\frac{d\kappa_t^2}{\kappa_t^2}\frac{\alpha_s(\kappa_t^2)}{2\pi}\int_{k_0^2}^{\kappa_t^2}\frac{dk_t^2}{k_t^2}
\int_{x}^{\frac{Q}{(Q+k_t)}}dz\nonumber\\ &\Bigg[& f_c
\left(\frac{x}{z} , k_t^2,Q^2\right)+f_{\overline{c}}
\left(\frac{x}{z} , k_t^2,Q^2\right) \Bigg] P_{qq}(z)
 \label{eq:d}.
\end{eqnarray}
It should be noted that the above relations for the subprocess $g
\rightarrow q \overline{q}$ and $q \rightarrow qg$ are true only for
the region of the $pQCD$. But since we are working in the small $x$
region or the equivalently, the high energy, the contribution of the
non-$perturbative$ region can be neglected, and the dominant
mechanism of the proton $c$-quark electroproduction  is the
photon-gluon fusion (i.e. the subprocess $g \rightarrow q
\overline{q}$).

Finally, the structure function $F_{2}^{c\overline{c}}(x, Q^2)$ can
be calculated  by the sum of gluons,  the equation (\ref{eq:2}),
and quarks,  the equation (\ref{eq:d}), contributions.
\section{Results, discussions and  conclusions}
As we pointed out before, the present work aim is to study
the charm content of a proton in the frameworks of the $KMR$ and
$MRW$ approaches and validate these two formalisms. In this regard,
 the charm structure functions, i.e., the sum of
 ${F_2}^{c\overline{c}}_{g \rightarrow q \overline{q}}$ and
  ${F_2}^{c\overline{c}}_{q \rightarrow qg}$ of the equations (\ref{eq:2}) and
  (\ref{eq:d})
  are calculated
   by using the $UPDF$ of the $KMR$ and $MRW$ approaches,
   i.e., the equations (\ref{eq:8}), (\ref{eq:9}) and  (\ref{eq:10}), respectively.
In the panels (a) to (g) of figure 1, the charm structure
functions, ${F_2}^{c\overline{c}}(x, Q^2)$, are displayed, by using
the $KMR$ ( $KMR, MMHT2014-LO$, dash curves) and $MRW$
($MRW, MMHT2014-NLO$, full curves) approaches,
 as a function of $x$ for different values of $Q^2$=$ 6.5, 12, 25, 30, 80, 160 $ and $600$ $GeV^2$  with the input
  $MMHT2014$ set of $PDF$ (to generate the $UPDF$) at the $LO$ and
  $NLO$ approximations, respectively. These results are compared with the data given by the $ZEUS$
  collaboration \cite{4}
   and the $NLO$-$QCD$ $HERAPDF$ $1.5$ \cite{29} predictions based on the general-mass variable- flavour-number scheme
   ($GMVFNS$). As one should expect, the results of the $KMR, MMHT2014-LO$ and
   $MRW, MMHT2014-NLO$ approaches are very close to each other at the low hard scale ($Q^2$), but they become separated as the hard scale increases. On the other hand,
   they are very close to the experimental data, i.e., the $ZEUS$ (2014)
   data \cite{4} (the full circle points). As we stated above, the general mass variable flavour number scheme
    ($HERAPDF$ 1.5 $GMVFNS$, dash-dotted curve) $pQCD$ calculation
is also plotted for comparison. As it was noted in the reference
\cite{WattWZ},  we do not expect to get a better results than $pQCD$,
although the $k_t$-factorization is more computationally simplistic. It is worth noting that the discrepancy between the data and the $k_t$-factorization prediction can be reduced by refitting the input integrated $PDF$. As it has been explained  in the reference \cite{WattWZ}, this treatment is adequate for initial investigations and descriptions of exclusive processes. In the panel
(d) of this figure, a comparison is also made with  the saturation
model introduced by $Golec-Biernat$ and $W\ddot{u}sthoff$ \cite{GBW}
($GBW$, the dotted curve) and the old  $ZEUS$ (2000) data \cite{zeus}
(the filled squares). Again our calculations are consistent with
them.

The charm structure functions, ${F_2}^{c\overline{c}}(x, Q^2)$, as a
function of the hard scale $Q^2$ are also calculated in the $KMR$
(dash curves), and $MRW$ (full curves) approaches as a function of
$Q^2$ for various $x$ values through the set of $MMHT2014$ $PDF$ as
the inputs. In the figure 2,
 the obtained results are compared with the data given by the $H1$ collaboration \cite{21} (full circles) and the $GMVFNS$ $QCD$
 predictions \cite{21-1,21-2,21-3,21-4,21-5} of $MSTW$  at $NNLO$ \cite{3} (dash-dotted
 curves).  Again as $Q^2$ and $x$ increase, the $MRW$
 prescription gives closer results with respect to those of $KMR$
 and the differences become larger. On the other hand, one can
 conclude that in general, the $k_t$-factorization and $pQCD$
 calculations are very closed and they are in agreement with the data.
 \\Also, we calculate the charm structure functions, ${F_2}^{c\overline{c}}(x, Q^2)$ by using the $LO$ collinear factorization with the inputs $MMHT2014$-$LO$ $PDF$ and plot the results in the figures 1 and 2. As expected, at higher energies ($Q^2$), the compatibility between the $k_t$ and collinear factorization calculations with the same $PDF$ becomes greater.
 \\To make the comparison more apparent between the frameworks of $KMR$ and $MRW$, the typical input of the gluon and charm quark $PDF$ at
scale $Q^2 = 25 GeV^2$, by using the  $MMHT2014$-$LO$  (dash curves) and
$MMHT2014$-$NLO$ \cite{22} (full curves), are plotted in the figure 3, and  the $KMR$-$UPDF$ (dash curves) and $MRW$-$UPDF$ (full curves) are plotted versus $k_t^2$
at typical values
of $x$ = 0.1, 0.01 and 0.001 and the factorization scale
$Q^2 = 100 GeV^2$ in the figure 4.

As shown in the figure 4 for the large values of $x$ (see the panel (d)), the values of  the $KMR$-$UPDF$ are larger than  the $MRW$-$UPDF$ due to increasing the scale $k^2=\frac{k_t^2}{1-z}$ relative to the scale $k_t^2$. 
This increase in scale, reduces $\alpha_s$ and $PDF$ and consequently decreases the $MRW$-$UPDF$ relative 
to the $KMR$-$UPDF$. But in the panel (a), since the charm quark $PDF$ of the $MMHT2014$-$NLO$ are larger than
 the charm quark $PDF$ of the $MMHT2014$-$LO$  (see the figure 3), the $MRW$-$UPDF$ increase relative to the $KMR$-$UPDF$ at the small $k_t^2$.
But for the small values of $x$ and the small $k_t^2$, the scales of the two approaches ($k_t^2$ and $k^2$) are almost equal, and the difference in the $KMR$-$UPDF$ and $MRW$-$UPDF$ is related to how the cutoff (due to $AOC$) is applied. As it is mentioned in section $II$, unlike the $KMR$ approach, in the $MRW$ approach, the effect of $\Delta$ on the terms which include non-singular splitting functions is negligible. Therefore, the $MRW$-$UPDF$ becomes larger than the $KMR$-$UPDF$ ones at the small $x$. This increase is more pronounced  for  the charm-$UPDF$ than the gluon-$UPDF$. In the explanation of this increase, it can be argued that, as shown in the figure 3, since at the small $x$, gluons are much larger than charm quarks, this increase should happen. Therefore, terms containing quarks in the equations (\ref{eq:8}), (\ref{eq:9}) and (\ref{eq:10}) can be ignored in comparison to those containing gluons. As a result, it is natural that both sets of the gluon-$UPDF$  become very similar, but due to the presence of non-singular terms, such as
 $P_{qg} \times g$ in quark-$UPDF$, the charm-$UPDF$ of the $MRW$ approach becomes larger than the charm-$UPDF$ of the $KMR$ approach. Here, it should be noted that, as it was shown in the references \cite{16,17,18,19}, the $KMR$ formalism suppresses the discrepancies between the inputs $PDF$. Therefore, although gluons are larger in the $LO$ approximation than in the $NLO$ approximation (see the figure 3), the $MRW$-$UPDF$ are still larger than the $KMR$-$UPDF$ (at the small $k_t^2$) because of the difference in the use of the cutoff $\Delta$. 
For the small $k_t^2$,  due to the increase in $k ^ 2$ over $k_t^2$, the lower limit of integral increases in the Sudakov form factor  the equations (\ref{eq:5}) and (\ref{eq:12}), so power of the exponential function becomes smaller, and as a result the Sudakov form factor in the $MRW$ approach increase relative to the Sudakov form factor of the $KMR$ approach. Therefore, the  $MRW$-$UPDF$ becomes larger than the $KMR$-$UPDF$ at the small $k_t^2$, except for the large $x$ region due to the decrease in the $\alpha_s$ and the $PDF$  in the large scale $k^2$. 
For large $k_t^2$,  due to the presence of the cutoff  $k^2<\mu^2$ in the $MRW$ approach, the $MRW$-$UPDF$ are smaller than $KMR$-$UPDF$. 
Given the structure function equation ($F_2$), since it is proportional to the expression $\frac{\alpha_s(\mu^2)}{k_t^2}$, so it is clear that the contribution of small $k_t^2$ is dominant. Therefore, given that the $MRW$-$UPDF$ are larger than the $KMR$-$UPDF$ in the small $k_t^2$, the charm structure functions which are extracted from the $MRW$ approach is larger than those of $KMR$. As we expected, and it is clear from figures 1 and 2, the difference is more significant in the larger $Q^2$. 
As the figures 1 and 2 illustrate and we expected, since charm quark is mostly produced at high energies and therefore at small $x$, the use of the $UPDF$ in the $NLO$ approximation ($MRW$-$UPDF$) can be in better agreement with the experimental data.

In conclusion, we can conclude that the obtained results for the
charm structure functions with the predictions of
 the $k_t$-factorization formalism
by using the $unintegrated$ parton distribution functions ($UPDF$),
which are generated through the $KMR$ and $MRW$ procedures are in agreement with the
predictions of the $pQCD$ and the experimental data .
 But the charm structure functions, which are extracted from the $MRW$ approach,
  have a better agreement to the experimental data with respect to that of $KMR$.
   In explaining the cause of this phenomena, we can conclude  that: This happens because  the $unintegrated$ parton
    distribution functions of the $MRW$ approach (at the small $x$    and  small $k^2_ t$ regions,
     i.e., the $c$-quark production domain) are slightly larger than the $unintegrated$ parton distribution
     functions of the $KMR$ formalism \cite{10}, which is due to the use the scale $k^2$ instead of the scale $k_t^2$ and not imposing $AOC$ constraint on non-singular terms in the $MRW$-$UPDF$. 
\begin{acknowledgements}
$NO$ would like to acknowledge the University of $Bu-Ali Sina$ and
Dr. $M.$  $Hajivaliei$ for their support. $MM$ would also like to
acknowledge the Research Council of University of Tehran   for the
grants provided for him.
\end{acknowledgements}

\newpage
\begin{table}[ht]
\includegraphics[width=15cm, height=15cm]{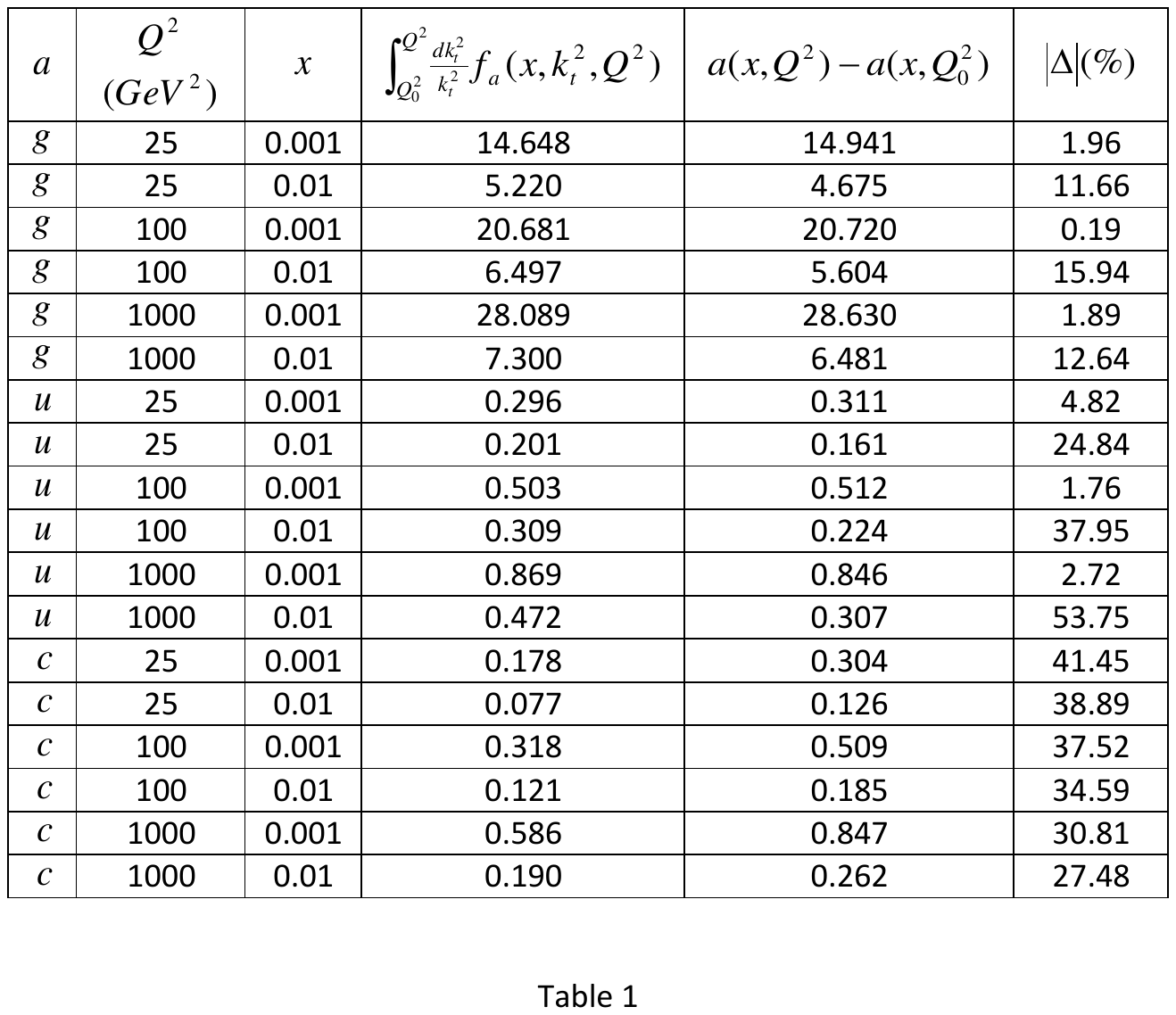}
\caption{The comparison of the $k_t$-integrals of the $KMR$-$UPDF$, by using $MMHT2014-LO$ $PDF$ as the inputs, and their corresponding ordinary $PDF$. }
\end{table}

\begin{figure}[ht]
\includegraphics[width=15cm, height=15cm]{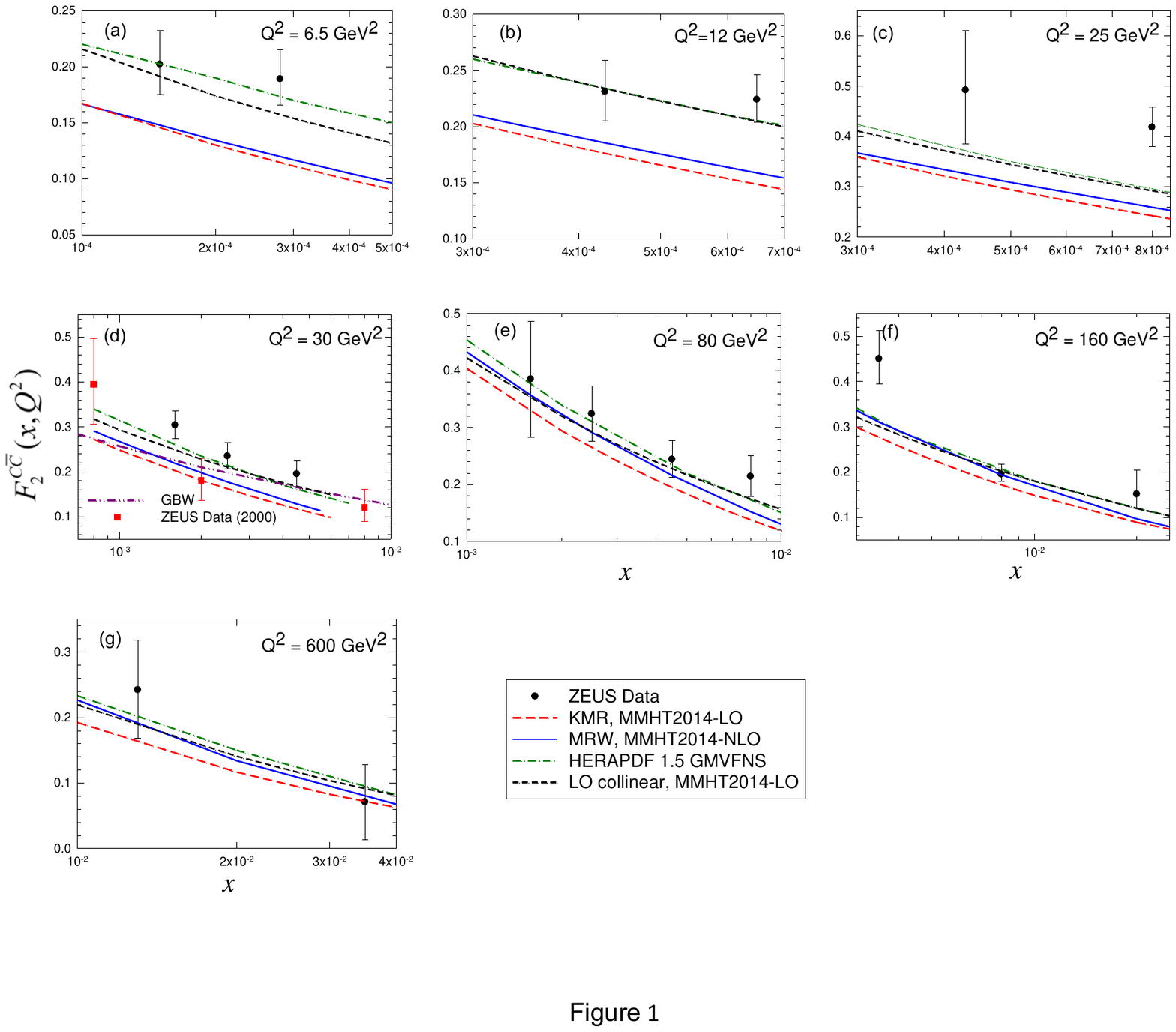}
\caption{The structure functions $F_{2}^{c\overline{c}}(x, Q^2)$
  as a function of $x$ for various $Q^2$ values for two different schemes, namely $KMR$ and $MRW$. The $NLO$ $QCD$ $HERAPDF$ $1.5$
  \cite{29}, $GBW$ \cite{GBW}
 predictions and $ZEUS$ data \cite{4,zeus} are also given. See the text for more explanations.}
 \label{fig:1}
\end{figure}
\begin{figure}[ht]
\includegraphics[width=15cm, height=15cm]{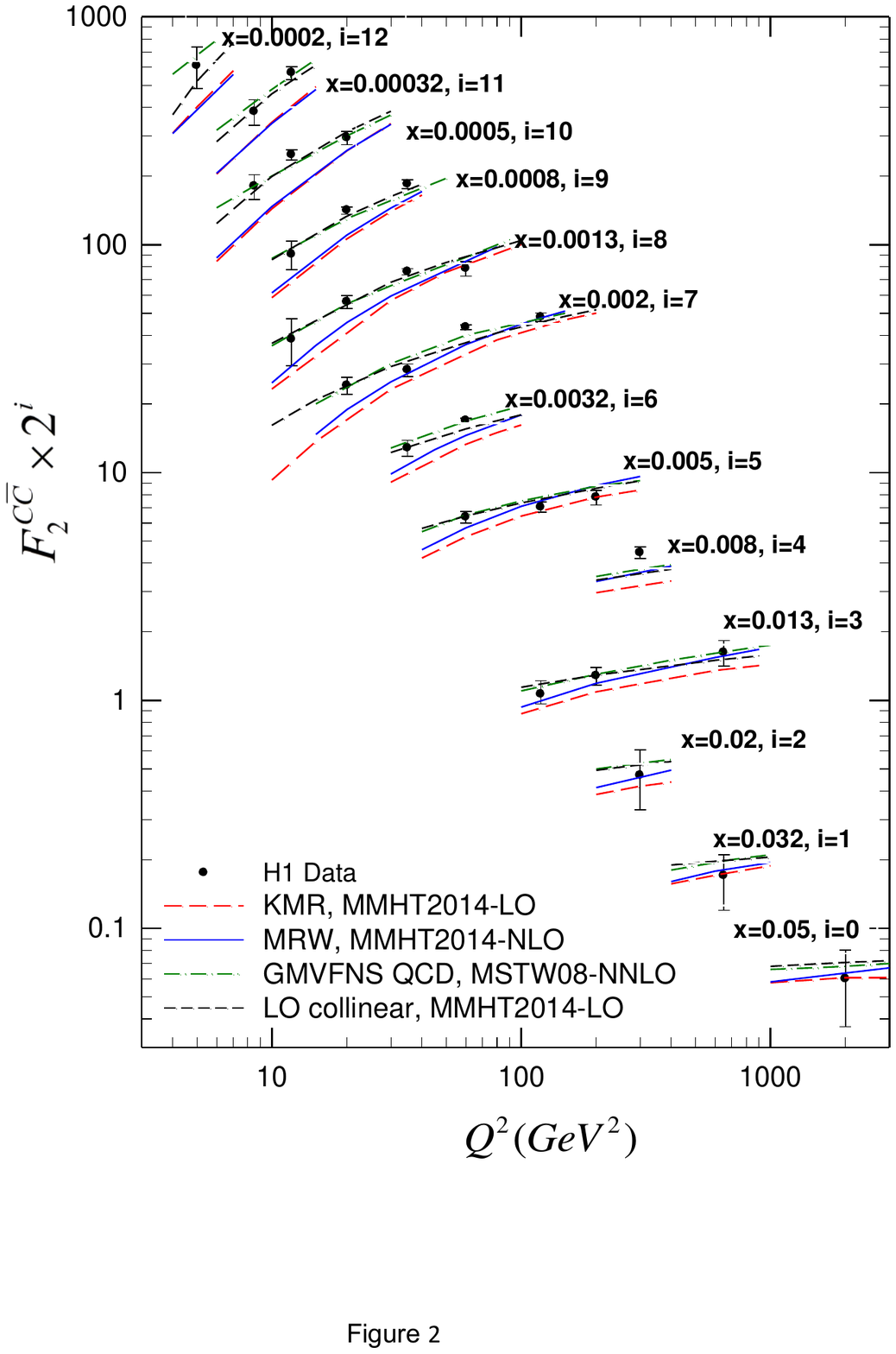}
\caption{The comparison of the charm structure functions, in the
frameworks of $KMR$ and $MRW$ by using the $MMHT2014-LO$ and
$MMHT2014-NLO$ $PDF$ data \cite{22}, respectively, as a function of
$Q^2$ for various $x$ values, with the $GMVFNS$ $QCD$ predictions \cite{21-1,21-2,21-3,21-4,21-5} of
$MSTW$ at $NNLO$ \cite{3} and $H1$ data \cite{21}.}
\label{fig:2}
\end{figure}
\begin{figure}[ht]
\includegraphics[width=15cm, height=15cm]{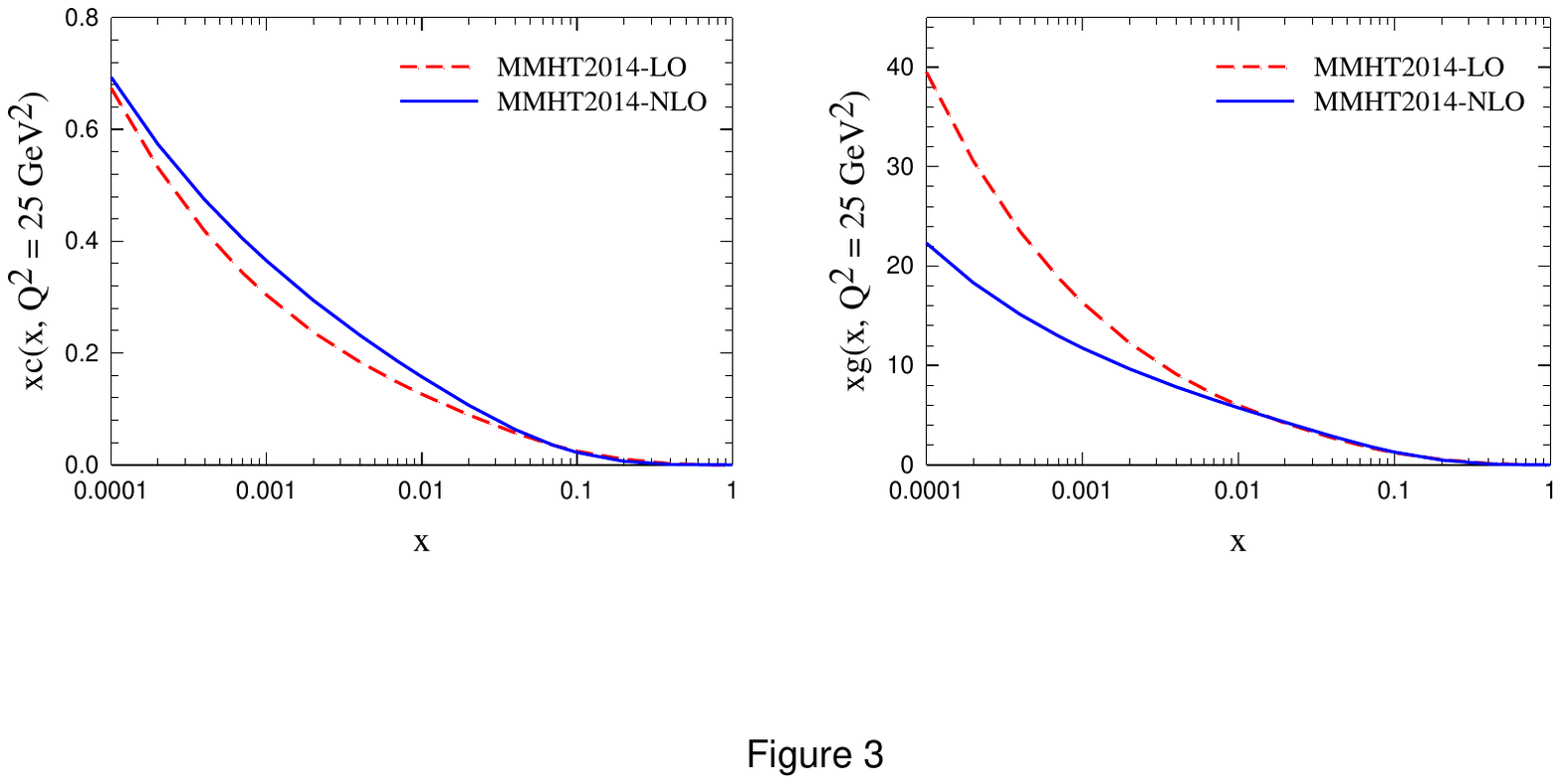}
\caption{The integrated charm quark  and gluon distribution functions at
scale $Q^2=25 GeV^2$, by using the $MMHT2014$-$LO$ and $MMHT2014$-$NLO$ $PDF$ data \cite{22}.}
\label{fig:3}
\end{figure}
\begin{figure}[ht]
\includegraphics[width=15cm, height=15cm]{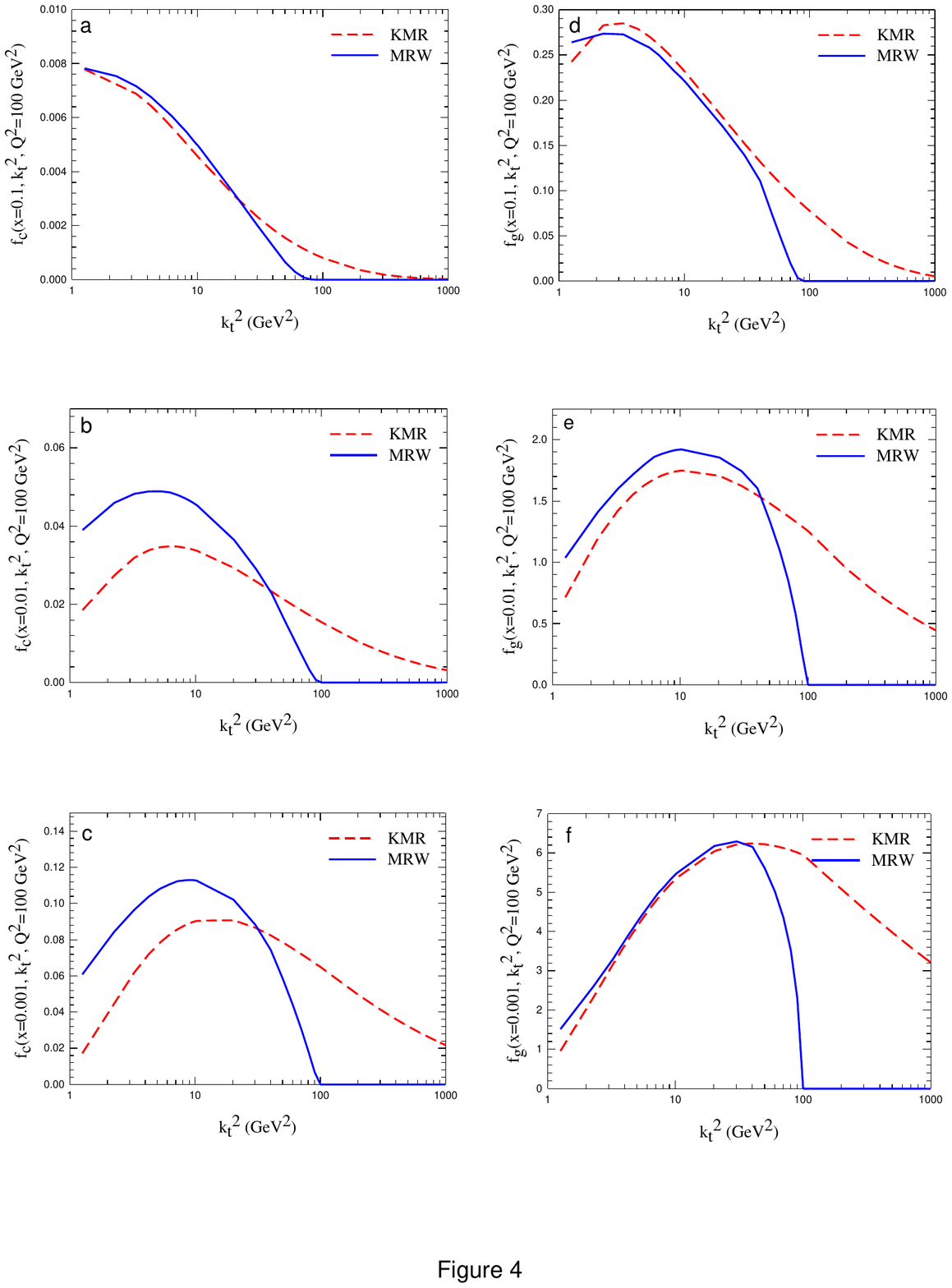}
\caption{The unintegrated charm quark and  gluon distribution functions versus $k_t^2$ with the $KMR$ ($MRW$)
prescriptions by using the $MMHT2014-LO$ $PDF$ ($MMHT2014-NLO$ $PDF$) as the inputs.}
\label{fig:4}
\end{figure}
\end{document}